\documentclass[12pt]{amsart}
\usepackage{amsmath, amssymb, amsthm, amsfonts, mathrsfs}
\usepackage{cite}
\usepackage{amscd}
\usepackage{url}
\usepackage{graphicx}
\usepackage{caption}
\usepackage{subcaption}
\usepackage{fullpage}
\usepackage{xspace}

\theoremstyle{definition}

\newcommand{\ip}{\ensuremath{\text{IP}_0}\xspace}
\newcommand{\zdx}{\ensuremath{\text{ZD}_{\chi}}\xspace}
\newcommand{\zdr}{\ensuremath{\text{ZDR}}\xspace}
\newcommand{\zdt}{\ensuremath{\text{ZD}_t}\xspace}
\newcommand{\mars}{\ensuremath{\text{MaRS}^{*}}\xspace}

\makeatletter
\g@addto@macro{\endabstract}{\@setabstract}
\newcommand{\authorfootnotes}{\renewcommand\thefootnote{\@fnsymbol\c@footnote}}%
\makeatother

\begin{document}

\date{}

\begin{center}
  \LARGE 
  The Art of War: Beyond Memory-one Strategies in Population Games \par \bigskip

  \normalsize
  \authorfootnotes
  Christopher Lee\footnote{corresponding author, email: leec@chem.ucla.edu}\textsuperscript{1,2,3}, Marc Harper\textsuperscript{3}, Dashiell Fryer\textsuperscript{4}\par \bigskip

  \textsuperscript{1}Department of Chemistry \& Biochemistry, UCLA, \par
  \textsuperscript{2}Department of Computer Science, UCLA, \par
  \textsuperscript{3}Institute for Genomics and Proteomics, UCLA, \par
  \textsuperscript{4}Department of Mathematics, Pomona College\par \bigskip
  
\end{center}

\begin{abstract}
We define a new strategy for population games based on techniques from machine learning and statistical inference that is essentially uninvadable and can successfully invade (significantly more likely than a neutral mutant) essentially all known memory-one strategies for the prisoner's dilemma and other population games, including ALLC (always cooperate), ALLD (always defect), tit-for-tat (TFT) \cite{imhof2007tit}, win-stay-lose-shift (WSLS) \cite{nowak1993strategy}, and zero determinant (ZD) strategies, including extortionate and generous strategies. We will refer to a player using this strategy as an \emph{information player} and the specific implementation as \ip. Such players use the history of play to identify opponent's strategies and respond accordingly, and naturally learn to cooperate with each other.
\end{abstract}

\section{Introduction}
The prisoner's dilemma (PD) \cite{rapoport1965prisoner} has a long history of study in evolutionary game theory \cite{nowak1990stochastic} \cite{nowak2006evolutionary} and finite populations and is usually defined by a game matrix
$ \left( \begin{smallmatrix}
R & S \\
T & P
\end{smallmatrix}\right) $
with $T > R > P > S$ and often $2R > T+S$. A special case known as the \emph{donation game} is given by $T=b$, $R=b-c$, $P=0$, $S=-c$, with $0 < c < b$. 
The discovery of zero determinant strategies by Press and Dyson \cite{press2012iterated} has invigorated the study of the prisoner's dilemma, including the evolutionary stability of these strategies in population games and their relationship to and impact on the evolution of cooperation \cite{nowak1990stochastic} \cite{stewart2012extortion} \cite{hilbe2013adaptive} \cite{hilbe2013evolution} \cite{akin2012stable} \cite{adami2013evolutionary} \cite{roemheld2013evolutionary}. In a tournament emulating the influential contest conducted by Axelrod \cite{axelrod2006evolution}, Stewart and Plotkin show that some zero determinant (ZD) strategies are very successful; Adami and Hintze \cite{adami2013evolutionary} have shown that ZD strategies are evolutionarily unstable in general, but can be effective if opponents can be identified and play can depend on the opponent's type (including versus itself). In particular, how a strategy fares against itself becomes crucial in population games.

Many strategies for the prisoner's dilemma have been studied in a huge array of contexts,
and it is often found that simpler strategies can beat
more complex strategies (e.g. TFT won early repeated prisoner's dilemma
tournaments \cite{axelrod2006evolution}).  
It has long been common to formulate PD strategies
as first-order Markov processes, i.e. 
whose next move depends only on the last game outcome.
This can be described by a vector of four probabilities denoting 
the probability that the player will select to cooperate (C) 
based on the previous round of play: 
$(Pr(C | CC), Pr(C | CD), Pr(C | DC), Pr(C | DD))$;
we will refer to this as a {\em strategy vector}.
Press and Dyson suggested that such first-order Markov strategies,
called memory-one strategies, 
can dominate more complex strategies; specifically, that using 
higher-order history does not help versus a ZD (first-order Markov)
strategy \cite{press2012iterated}.  Stewart and Plotkin have also
argued that a generous ZD strategy can be robust against {\em any}
invading strategy (i.e. no invader can achieve better than neutral 
fixation probability) \cite{stewart2012extortion} under a set of assumptions
including weak selection. In population games, Adami and Hintze have indicated 
that {\em tag} information identifying
which players are of the opposing type can significantly
increase evolutionary success \cite{adami2013evolutionary}.  
They also suggested that it is possible to recognize an opponent's 
strategy from the history of play. Can information from past history 
-- ignored by a first-order Markov strategy --
improve evolutionary success?

\subsection{Information Players}
We refer to a player that uses such information (history or some sort of
tag indicating strategy) as an 
{\em information player} (IP).  Formally, whereas a first-order Markov
player's next move is conditionally independent of past history
given the current game outcome (zero mutual information),
an information player's next move depends on past history
(shares non-zero mutual information given the current game outcome).
Specifically, we investigate whether machine learning can yield
useful information from past history, both to identify opponents
and to infer their likely behavior.

Our approach recapitulates long-standing principles,
for example as summarized by Sun Tzu's \emph{The Art of War}:

\begin{quote}
\noindent The general who wins the battle makes many calculations in his temple before
the battle is fought. The general who loses makes but few calculations beforehand. \\

\noindent Know your enemy and know yourself, find naught in fear for 100 battles. \\

\noindent ...what is of supreme importance in war is to attack the enemy's strategy. \\

\noindent One defends when his strength is inadequate, he attacks when it is abundant.

-- Sun Tzu, \emph{The Art of War}
\end{quote}

In particular, we explicitly define a strategy that utilizes the history of play to
determine the strategies of other players (assuming no strategy identifying tag is supplied),
and uses these determinations to optimize its subsequent moves. We call this specific
implementation of an information player \ip, which embodies the principles above as follows:

\begin{itemize}
\item \emph{Know your enemy}. Rather than seeking to maximize its
{\em score}, \ip initially seeks to maximize its {\em information}
about another player's strategy vector. For the first 10 rounds vs.
a specific player, \ip selects its plays, either cooperate (C) or defect (D),
solely to maximize its information yield about the other player's 
strategy vector probabilities.  We refer to this as the 
\emph{information gain phase}. The four probabilities are estimated 
from these rounds of play and are continually refined in subsequent rounds.

\item \emph{Know yourself}. Each \ip individual attempts to identify whether 
each other other player is also \ip, based purely on
whether it appears to ``play like me'' (choose the same moves an \ip
would have chosen).  In particular, the information gain phase
produces a unique pattern of play, that can be quickly recognized (within 
3 - 10 moves), even in the presence of random noise (randomly flipped moves).
Note however that each \ip player acts completely independently;
different \ip in a population share no information and do not communicate.

\item \emph{Attack the enemy's strategy}. In subsequent rounds, each \ip 
seeks to maximize its own average score (and by extension that of all 
IPs in the population) vs. that of the opposing player type.
Specifically, it always seeks to cooperate with other \ip individuals; 
versus the opposing type,
it chooses the optimal strategy vector based on its estimate
of the opposing type's strategy vector. As rounds proceed, 
each \ip continues to update its estimate of opponents probabilities, 
and adjusts its play as needed to maximize its average score difference.

\item \emph{One defends... one attacks...} We will see that \ip 
naturally switches effective strategy depending on the proportion of 
\ip in the population, and the opponent strategy.
Commonly, \ip initially cooperates with the opposing type, 
when \ip is in the minority, and later defects
against the opposing type, when \ip is in the majority.
\end{itemize}



In this paper we test our approach on a variety of traditionally 
successful strategies and ZD strategies, but our results are not 
limited to such opponents, nor for that matter to the Prisoner's Dilemma
game. We also allow for errors in play, described by an ambient 
noise parameter $\epsilon$ since this provides a greater 
variety of strategic interactions between many classical 
players such as TFT and WSLS.

\subsection{ZD strategies}
Stewart and Plotkin have shown that for weak selection the class of 
generous zero determinant strategies is evolutionarily robust 
in the space of memory-one players and that these robust ZD strategies 
can invade other (extortionate) zero determinant strategies
\cite{stewart2012extortion}.  We will refer to these robust strategies
as ``\zdr'' throughout this paper, and will use as one key example
the \zdr($\chi=\frac{1}{2}$) case, which represents the best of
the class that Stewart and Plotkin designate as ``Good and robust''
\cite{stewart2012extortion}.  We will refer to extortionate
ZD strategies as ``\zdx''; see Methods for details.

\section{Results}

\subsection{One defends when his strength is inadequate, he attacks when it is abundant}

The long run evolutionary fitness of a player of type $I$ is determined by
its mean stationary score relative to that of players of the opposing group $G$, 
specifically
\begin{equation}
\overline{S_{I}}-\overline{S_{G}}=
\frac{m-1}{N-1}\overline{S_{II}}+\frac{N-m}{N-1}\overline{S_{IG}}
-\frac{m}{N-1}\overline{S_{GI}}-\frac{N-m-1}{N-1}\overline{S_{GG}}
\label{mean_stationary_score}
\end{equation}
where $m$ is the number of players of type $I$ in the population;
$N-m$ is the number of players of type $G$;
$\overline{S_{II}},\overline{S_{IG}},\overline{S_{GI}}, \overline{S_{GG}}$
are the average scores of players of type $I$ with each other vs. with a player
of the opposing group, and the opposing group player's average scores
with a player of type $I$ vs. with another group player.  
An optimal strategy for player $I$ is simply one that maximizes
$\overline{S_{I}}-\overline{S_{G}}$.  Note that this is
strongly dependent on the population fraction $f=m/N$; for small $f$
($m \ll N$), $\overline{S_{I}}-\overline{S_{G}}$ is dominated by
the $\overline{S_{IG}},\overline{S_{GG}}$ terms; whereas for large $f$
it is dominated by the $\overline{S_{II}},\overline{S_{GI}}$ terms.
Note that the two-player game considered by Press \& Dyson is a subcase
of this spectrum; specifically, the only case ($N=2$) where there is
only one possible value of $m$ ($m=1$) 
and $\overline{S_{I}}-\overline{S_{G}}$ reduces to the trivial form 
$\overline{S_{I}}-\overline{S_{G}}=\overline{S_{IG}}-\overline{S_{GI}}$. 
Note carefully that \ip attempts to maximize the stationary score difference, in
simulation we use the actual values from the prisoner's dilemma matrix, and do 
not assume that the payoffs are the stationary payoffs (in contrast to e.g. 
\cite{stewart2013extortion}).

\begin{figure}[h]
        \begin{subfigure}[b]{0.4\textwidth}
            \centering
            \includegraphics[width=\textwidth]{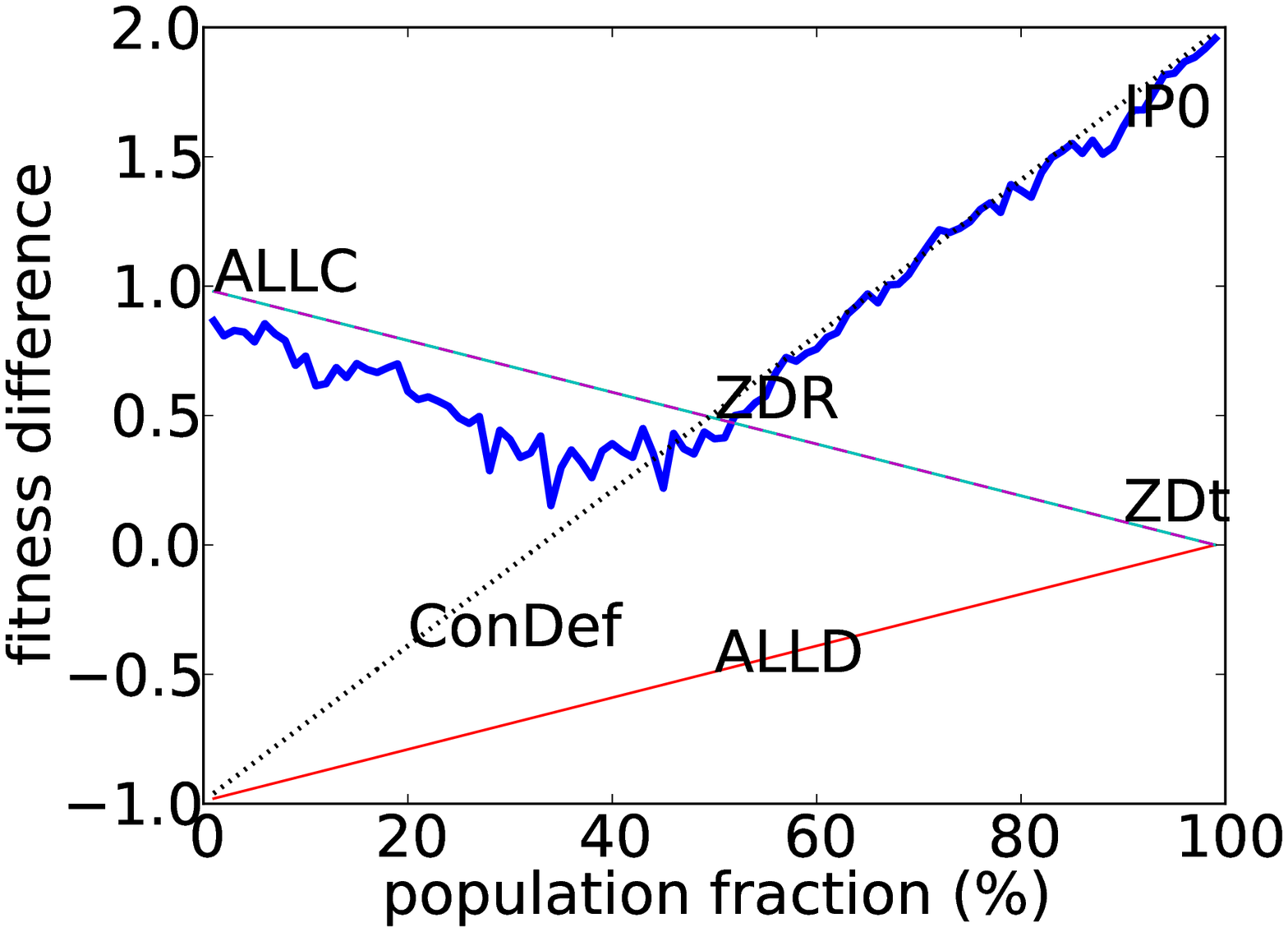}
            \caption{TFT, $\epsilon = 0$}
        \end{subfigure} \qquad
        \begin{subfigure}[b]{0.4\textwidth}
            \centering
            \includegraphics[width=\textwidth]{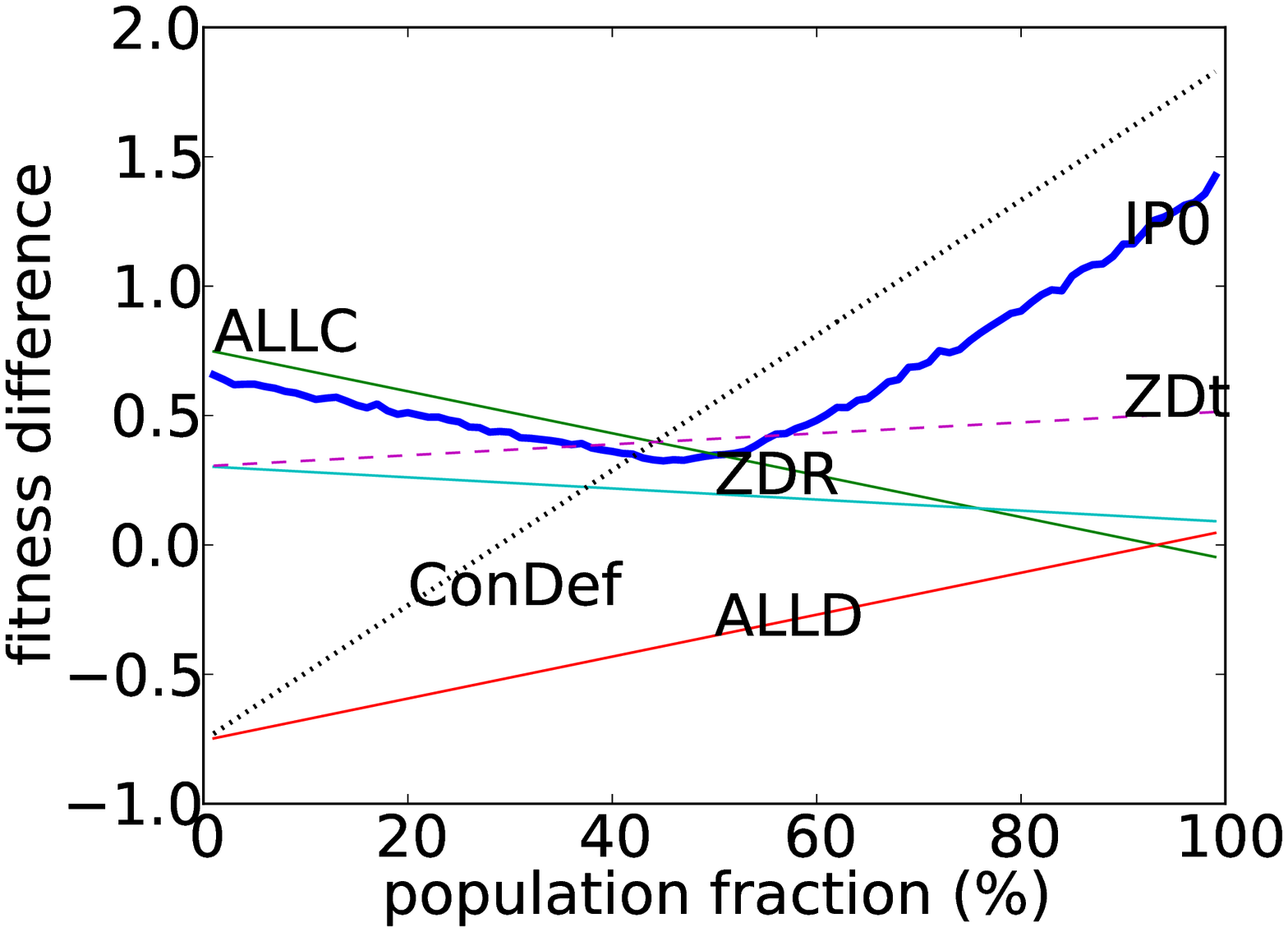}
            \caption{TFT, $\epsilon = 0.05$}
        \end{subfigure}%
        \\
        \begin{subfigure}[b]{0.4\textwidth}
            \centering
            \includegraphics[width=\textwidth]{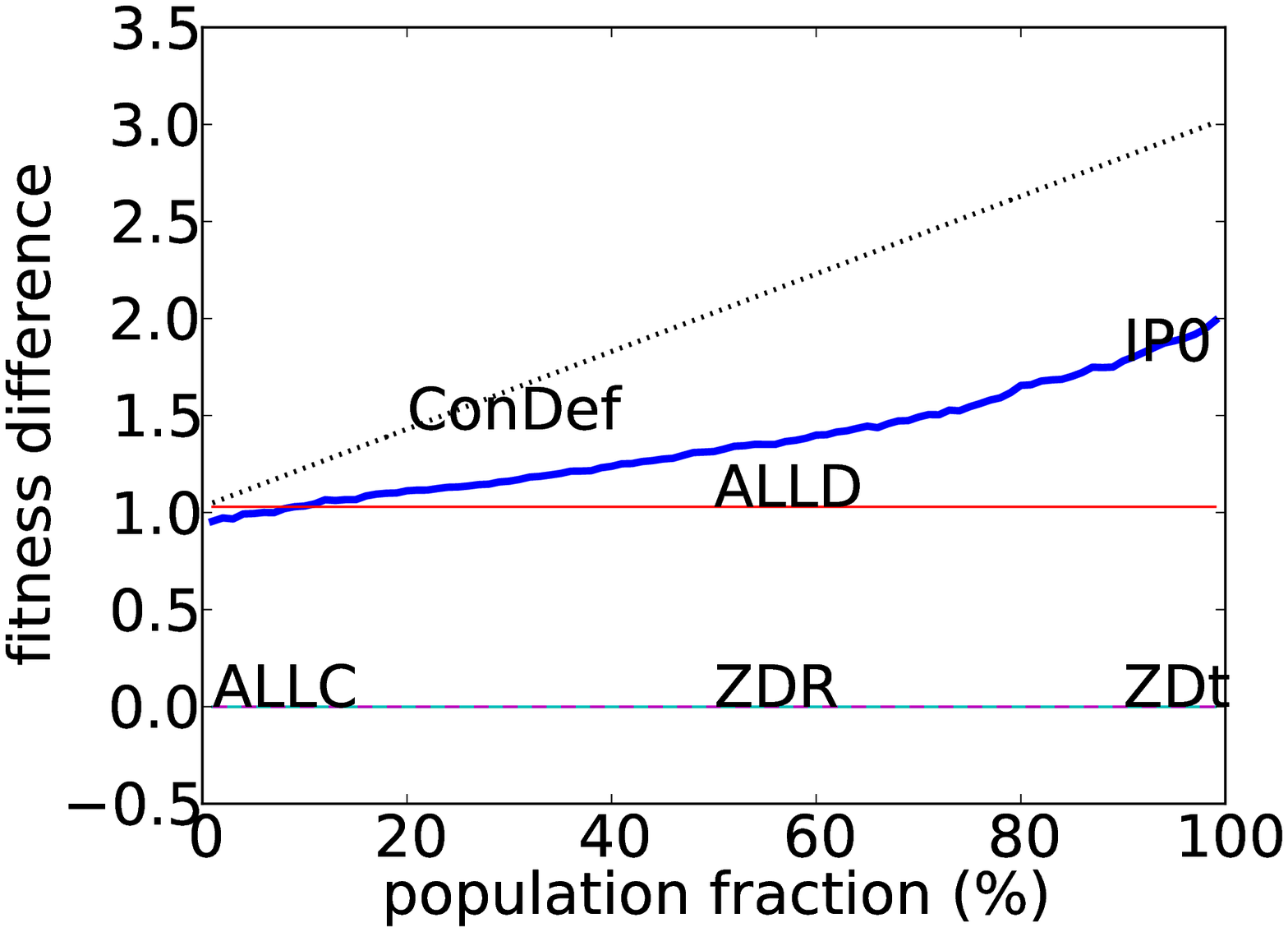}
            \caption{ALLC, $\epsilon = 0$}
        \end{subfigure} \qquad
        \begin{subfigure}[b]{0.4\textwidth}
            \centering
            \includegraphics[width=\textwidth]{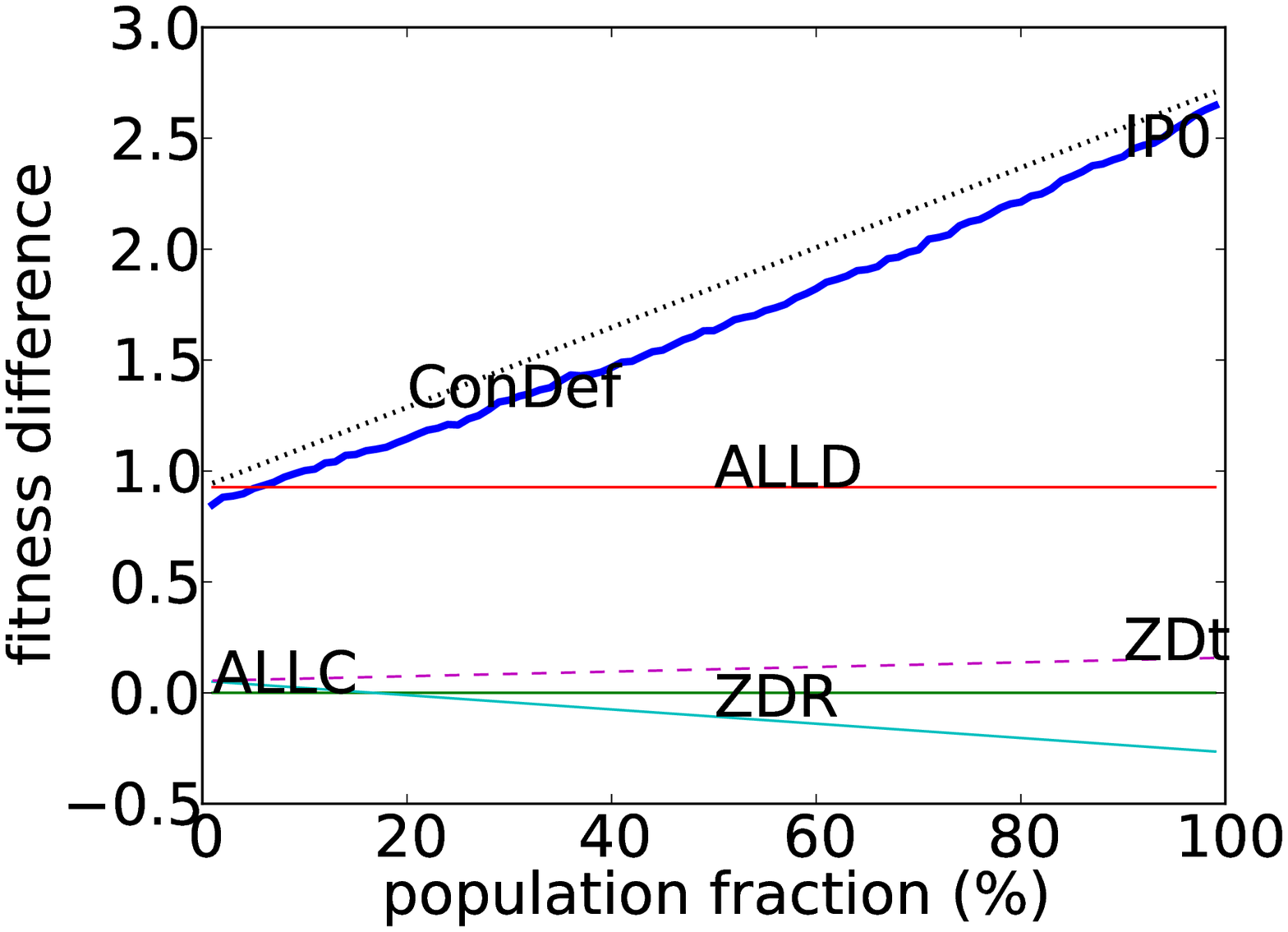}
            \caption{ALLC, $\epsilon = 0.05$}
        \end{subfigure}%
        \\
        \begin{subfigure}[b]{0.4\textwidth}
            \centering
            \includegraphics[width=\textwidth]{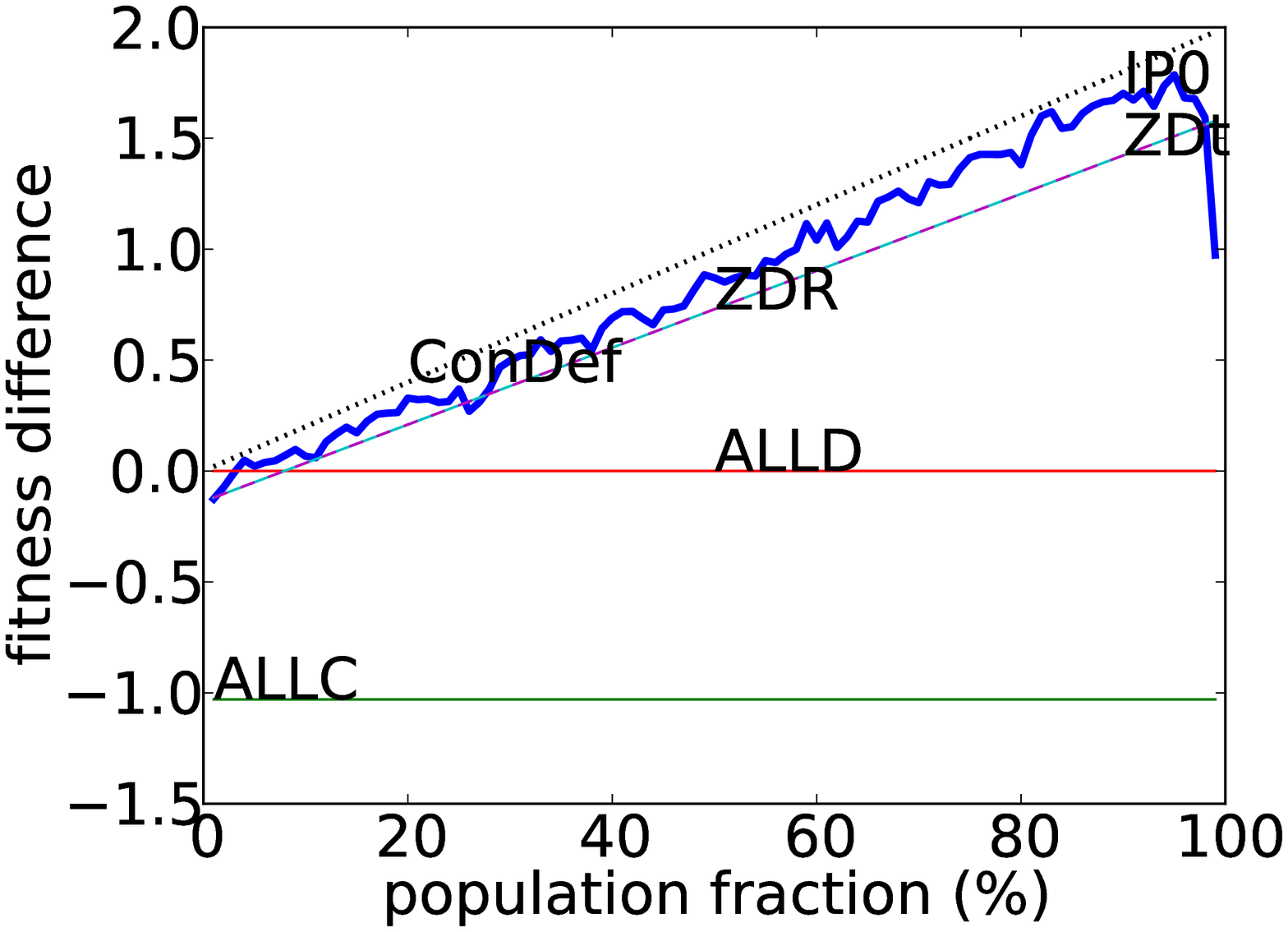}
            \caption{ALLD, $\epsilon = 0$}
        \end{subfigure} \qquad
        \begin{subfigure}[b]{0.4\textwidth}
            \centering
            \includegraphics[width=\textwidth]{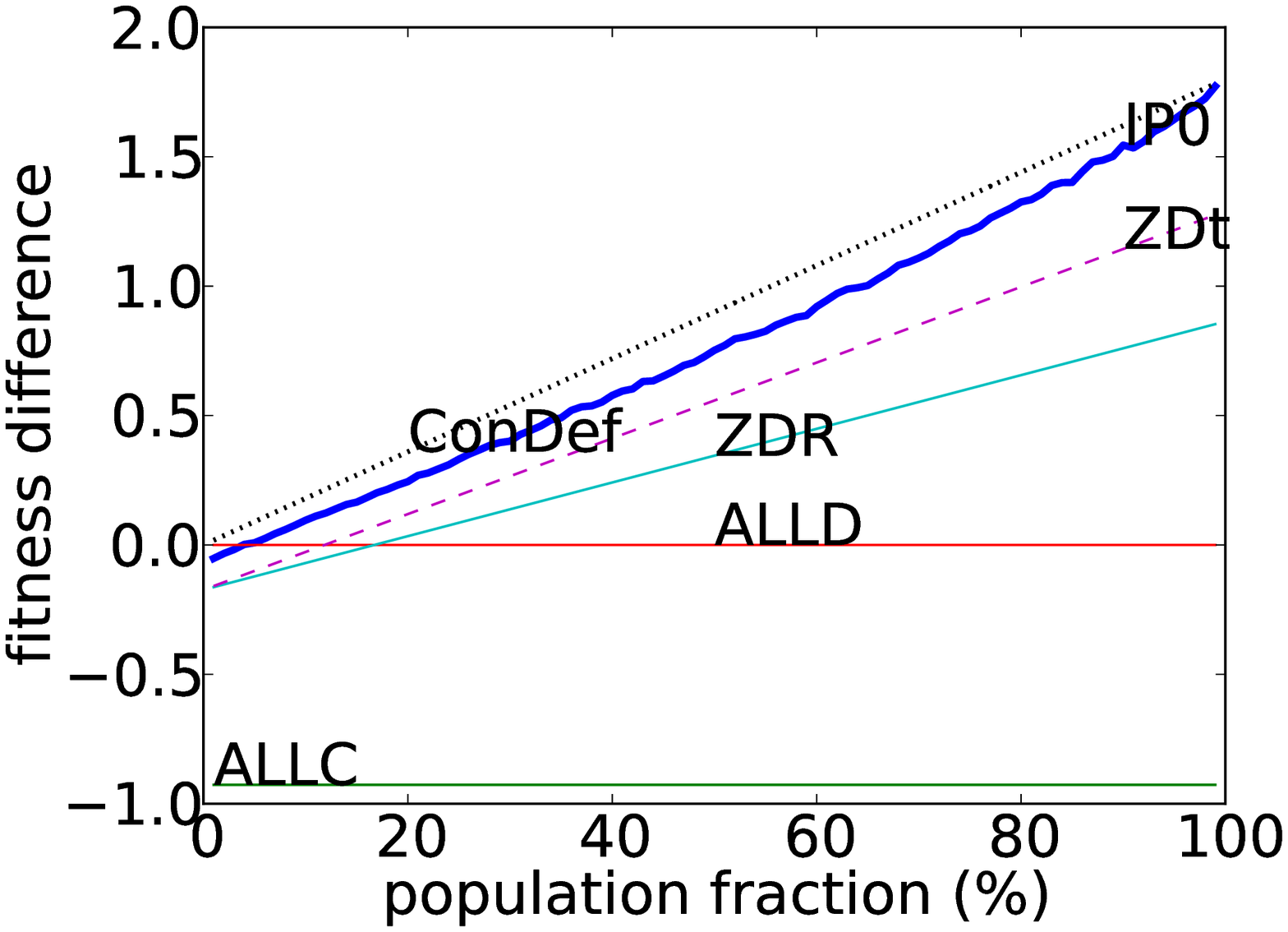}
            \caption{ALLD, $\epsilon = 0.05$}
        \end{subfigure}%
        \caption{Mean score difference $\overline{S_{I}}-\overline{S_{G}}$ as a function of the invader's population fraction, for several different invading strategies (I=ALLC, ALLD, \zdr, ConDef, \zdt) vs. several different resident strategies: TFT (A, B); ALLC (C, D); ALLD (E, F).}
        \label{figure_stationary_score_1}
\end{figure}

\begin{figure}[h]        
        \begin{subfigure}[b]{0.4\textwidth}
            \centering
            \includegraphics[width=\textwidth]{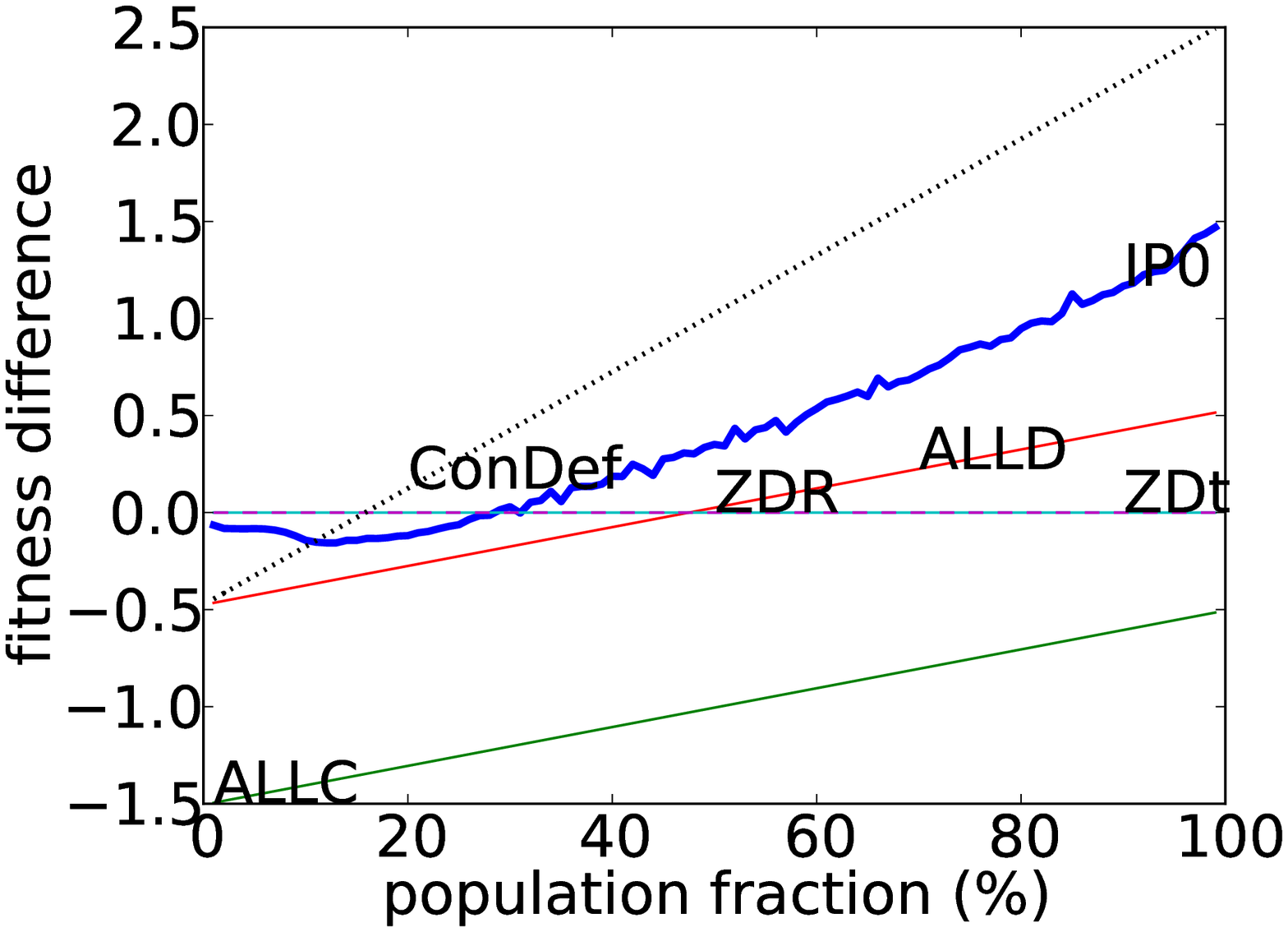}
            \caption{WSLS, $\epsilon = 0$}
        \end{subfigure} \qquad
        \begin{subfigure}[b]{0.4\textwidth}
            \centering
            \includegraphics[width=\textwidth]{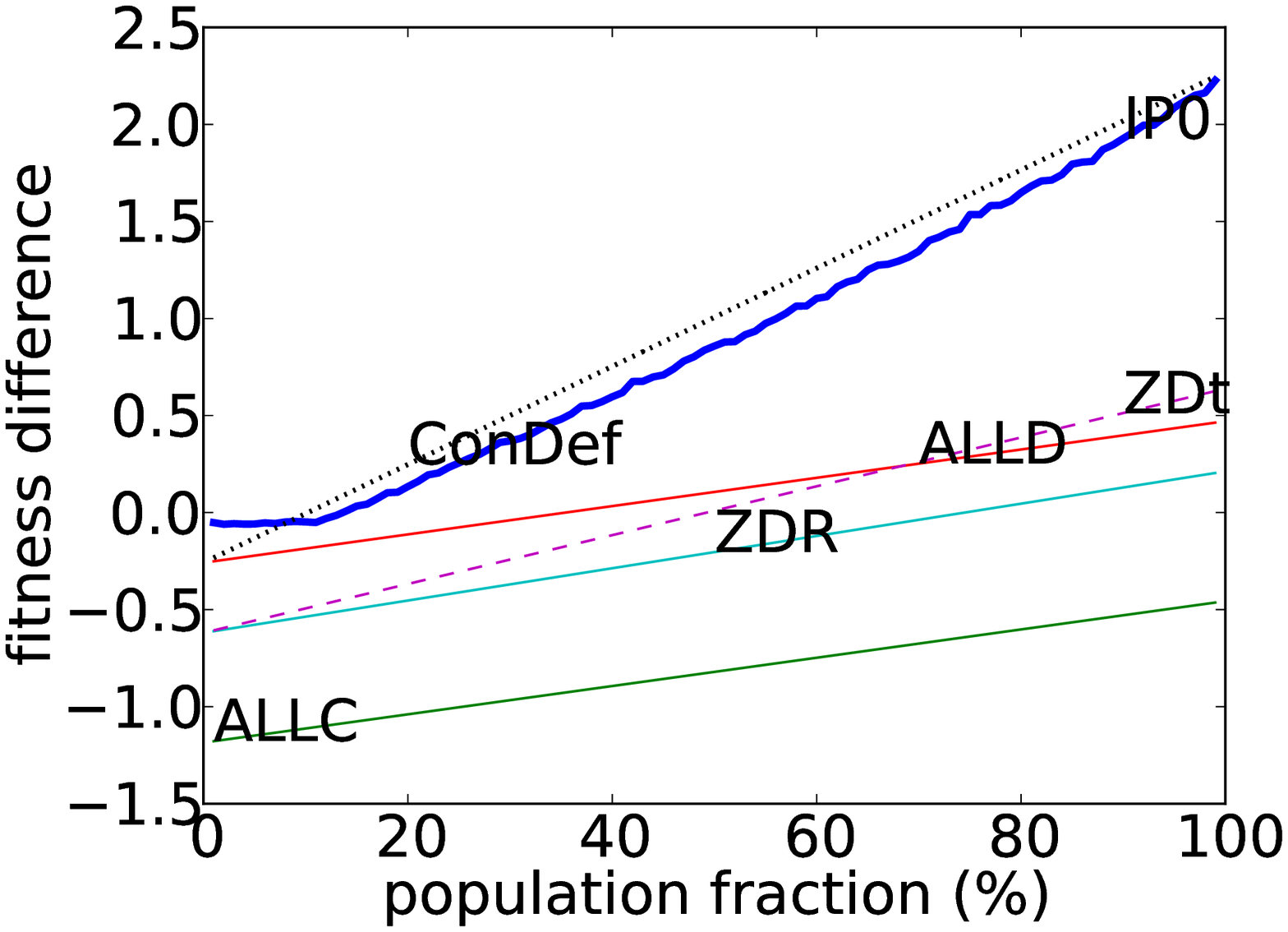}
            \caption{WSLS, $\epsilon = 0.05$}
        \end{subfigure}%
        \\
        \begin{subfigure}[b]{0.4\textwidth}
            \centering
            \includegraphics[width=\textwidth]{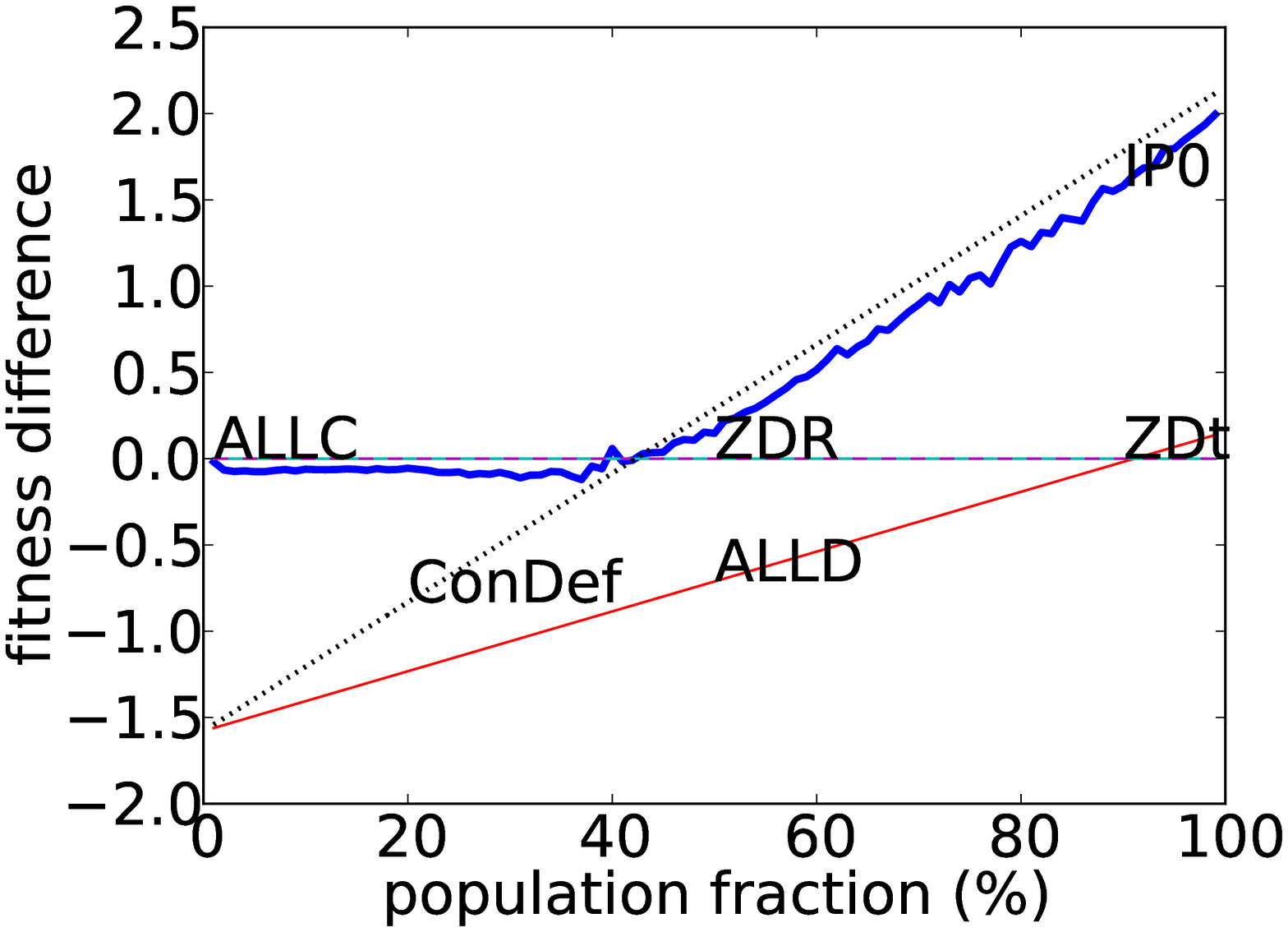}
            \caption{\zdr, $\epsilon = 0$}
        \end{subfigure} \qquad
        \begin{subfigure}[b]{0.4\textwidth}
            \centering
            \includegraphics[width=\textwidth]{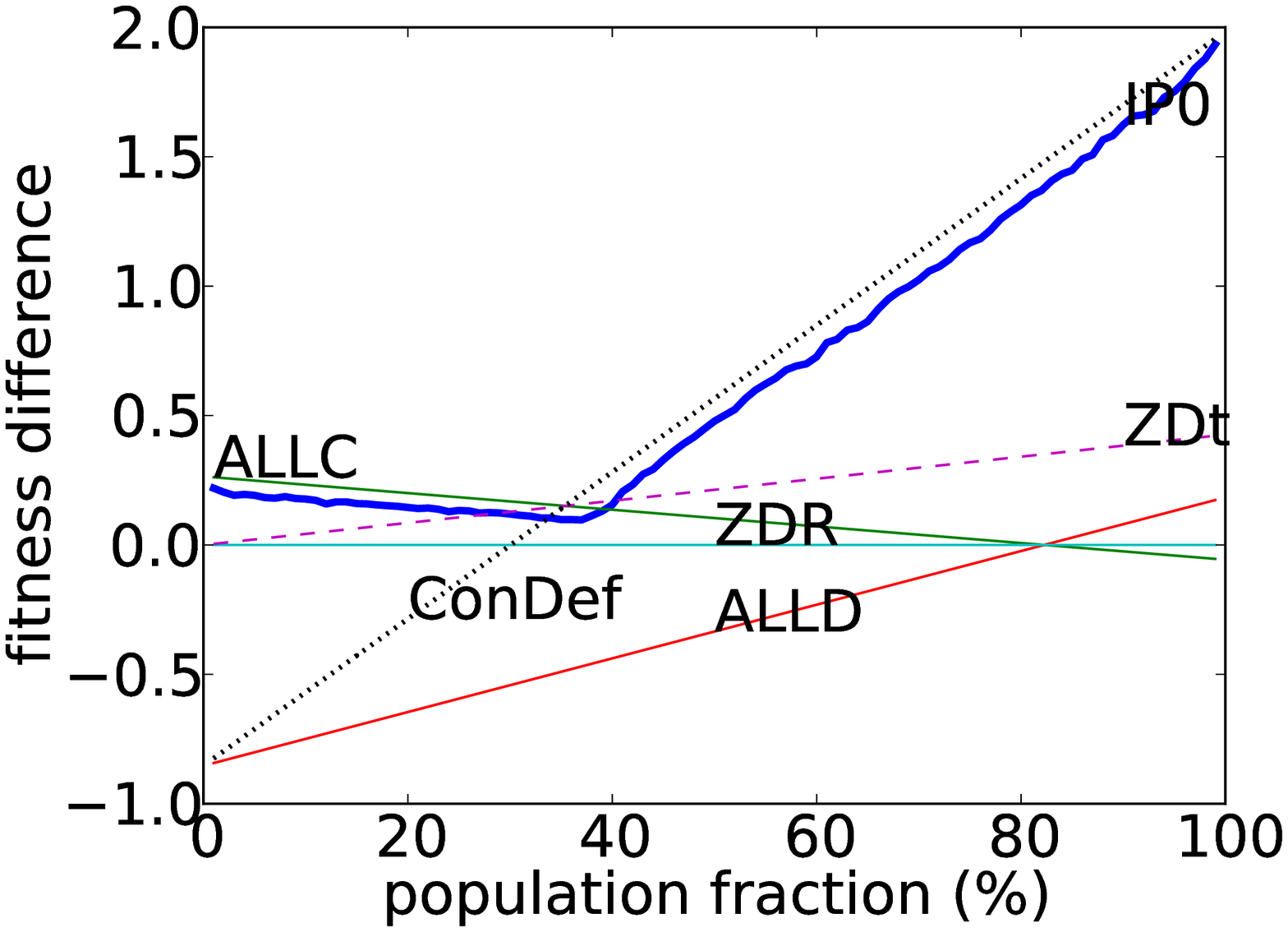}
            \caption{\zdr, $\epsilon = 0.05$}
        \end{subfigure}%
        \\
        \begin{subfigure}[b]{0.4\textwidth}
            \centering
            \includegraphics[width=\textwidth]{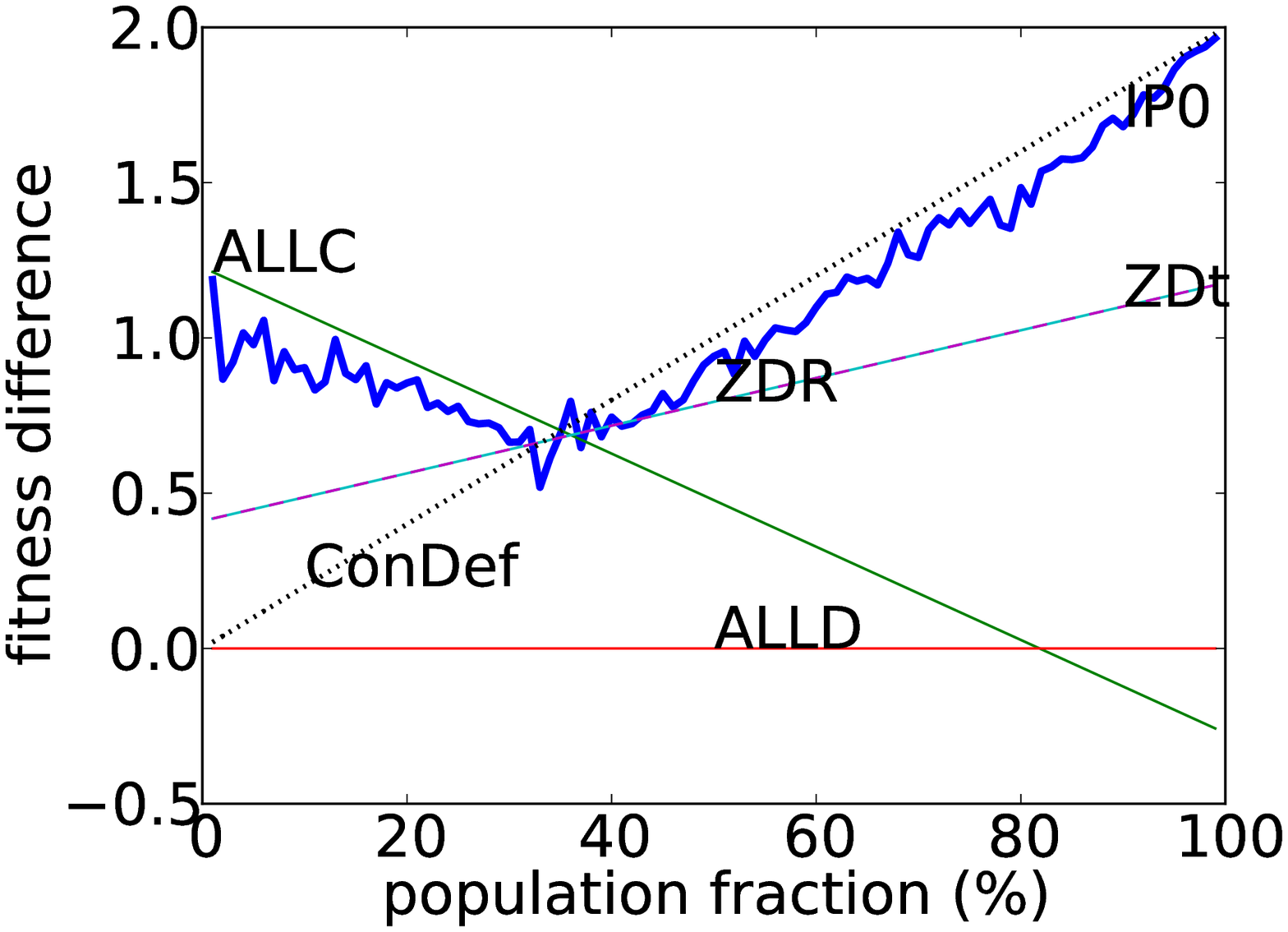}
            \caption{\zdx, $\epsilon = 0$}
        \end{subfigure} \qquad
        \begin{subfigure}[b]{0.4\textwidth}
            \centering
            \includegraphics[width=\textwidth]{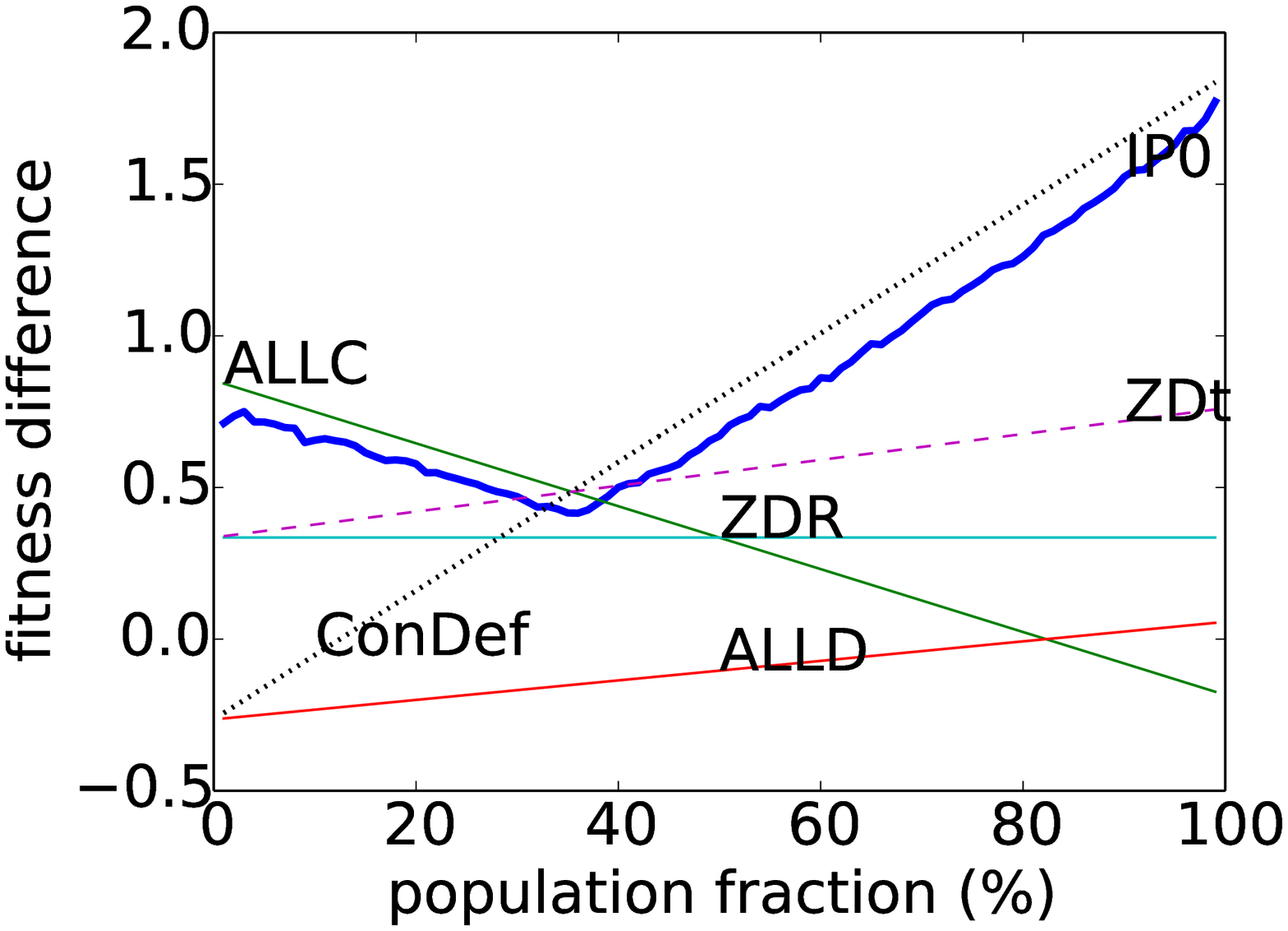}
            \caption{\zdx, $\epsilon = 0.05$}
        \end{subfigure}%
        \\

        \caption{Average fitness difference $\overline{S_{I}}-\overline{S_{G}}$ as a function of the invader's population fraction, for several different invading strategies (I=ALLC, ALLD,\zdr, ConDef, \zdt) vs. several different resident strategies: WSLS (A, B); \zdr (C, D); \zdx (E, F).}
        \label{figure_stationary_score_2}
\end{figure}

Figures \ref{figure_stationary_score_1} and \ref{figure_stationary_score_2} 
show $\overline{S_{I}}-\overline{S_{G}}$ as a function of population
fraction $f$, for a variety of established strategies, computed from
their long-term (stationary) scores \cite{press2012iterated}.
Several basic conclusions emerge from these plots.
First, no one strategy is optimal against all opponents: for example,
at low population fractions, \zdr is optimal against WSLS, whereas
ALLC is optimal against \zdx.  
Second, even against a single opponent, typically no one strategy is optimal
at all population fractions.  For example, against WSLS, \zdr scores better
than ALLD at low population fractions, but worse than ALLD at 
high population fractions.
Third, even at a single given point on such a score plot, it
is commonly not optimal for players of type $I$ to play the same strategy
vector with each other as with the opposing players of type $G$.
For example, at high population fractions, playing ALLD vs. the opponent
(ensuring $\overline{S_{GI}} \le P$) while playing ALLC with each
other (yielding $\overline{S_{II}}=R$) maximizes
$\overline{S_{I}}-\overline{S_{G}} \to R-P$.  Hintze and Adami have
posited a theoretical strategy, Conditional Defector (ConDef)
for achieving this: assuming that it is given the correct {\em tag}
for the type of each player, ConDef cooperates with other ConDef
players and defects vs. players of the opposing type
\cite{adami2013evolutionary}. (They also defined a tag-based ZD
player \zdt that cooperates with other \zdt players
and plays a ZD strategy against the opposing type).
Fourth, it is striking that even traditionally successful strategies such 
as WSLS and \zdr are vulnerable to invasion, because at low population
fractions an invader can achieve parity (neutral selection) vs. these
strategies, while at high population fraction it can gain a crushing
advantage over them (by switching to what is essentially ConDef).

Taken together, these results suggest that information gleaned from
the history of previous game outcomes can yield several basic
advantages for choosing moves in the subsequent rounds: player $I$ can
infer which individual players are ``like it'' (i.e. also of type $I$)
vs. ``enemy'' (i.e. of type $G$; we refer to this as {\em identification});
player $I$ can estimate player $G$'s strategy vector, 
enabling it to choose the optimal
counter-strategy; player $I$ can estimate what fraction of the
population consists of players of type $G$.  All of these are
crucial for maximizing $\overline{S_{I}}-\overline{S_{G}}$.


\subsection{Know Your Enemy and Know Thyself}

That identification is useful highlights a crucial question: in the absence of 
strategy-indicating tags,
can an information player determine the identity of other players 
($I$ vs. $G$) rapidly and accurately, purely from the past history of
how they played against it in previous games?
\ip begins its play versus
any other player with an {\em information gain} phase (infogain, see Methods).
This phase seeks to collect maximal information about the 
other's strategy vector, and at the same time estimates the
likelihood that the other is also an IP0 player; specifically
whether it is also playing by the infogain rule.  Thus the infogain
phase achieves self-recognition by a most basic principle, ``does the opponent
play similarly?'' (i.e. choose the same moves that \ip).

\begin{figure}[h!]
\includegraphics[width=0.4\textwidth]{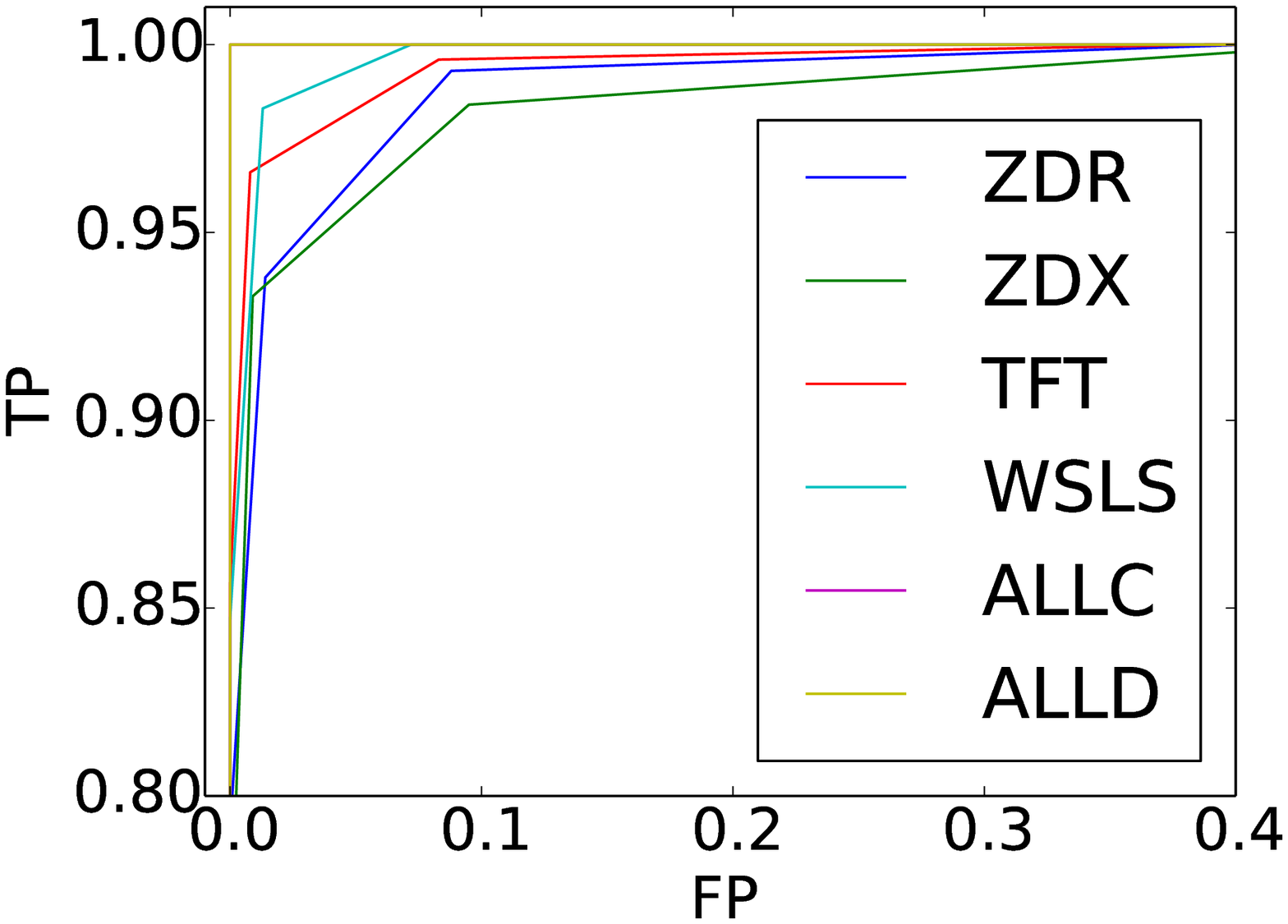}
\includegraphics[width=0.4\textwidth]{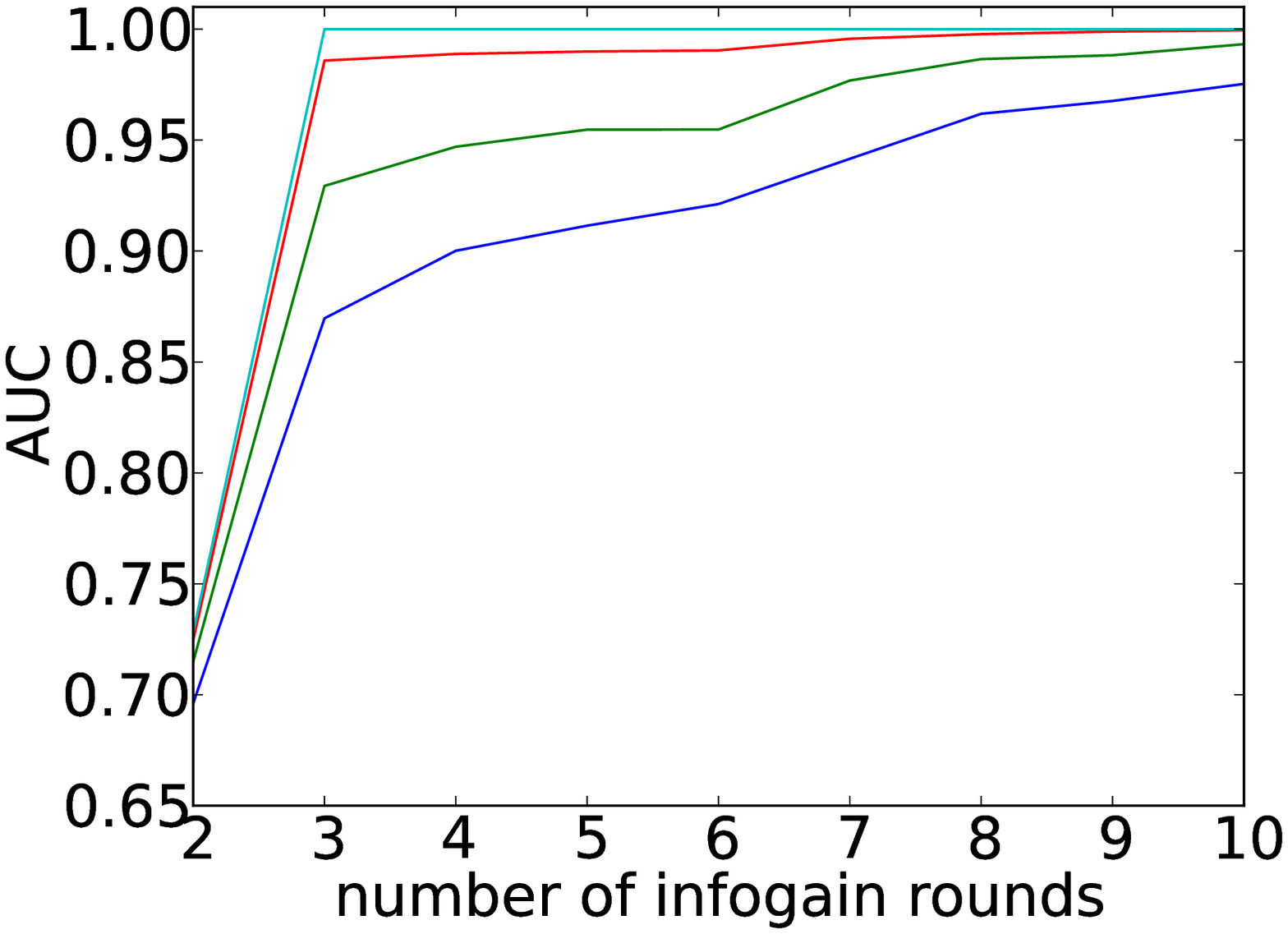}
\caption{Accuracy of information gain phase. Left: ROC for $\epsilon = 0.05$ and 10 infogain rounds. Vertical axis: true positives, Horizontal axis: false positives. \zdr is the hardest strategy to recognize among those tested. Right: AUC for IP against \zdr for $\epsilon = 0, 0.01, 0.05, 0.1$.}
\label{roc_auc}
\end{figure}

This approach can rapidly discriminate non-\ip players.  In the absence
of random noise (move errors), it is trivial: the very first move that
doesn't match the expected infogain move exposes the opponent to be
not \ip, and this typically occurs within the first 3 moves.
To make this more challenging, we assessed the effect of random
noise, by randomly flipping each player's move with probability $\epsilon$.
Then \ip must assess observed mismatches probabilistically, e.g.
by computing the probability that the observed mismatches could
have arisen from another \ip player due solely to random noise (see Methods).
This can achieve good discrimination, at the cost of a few extra rounds.
Figure \ref{roc_auc}A shows Receiver-Operator Characteristic (ROC)
curves for discrimination of non-\ip players 
(vertical axis, True Positives) vs.
IP0 players (horizontal axis, False Positives) 
after 10 rounds of play under 5\%
noise.  Corner strategies such as ALLC and ALLD were identified
perfectly (i.e. 100\% TP at 0\% FP), while the most difficult
case (\zdr) was identified with 98\% accuracy at a false positive rate
of only 10\%.  Concretely, for $N=100$ players, a
single \ip player invading a resident \zdr population could confidently
identify 97 of the 99 \zdr players, while having a 90\% probability
of recognizing any new \ip player within just 10 rounds of play.

To summarize the speed of this process and its sensitivity to noise,
we computed a standard measure of discrimination accuracy (AUC, Area
Under the Curve, the integral of the ROC curve) for the hardest case
(\zdr), and plotted it as a function of number of infogain rounds and
for different levels of noise (Figure \ref{roc_auc}B).  
At zero noise, perfect discrimination
(AUC=1) was achieved after just 3 rounds; with up to 10\% noise,
AUC accuracies of 87-98\% were attained after just 3 rounds.
Even at 10\% noise, AUC accuracy of greater than 
97\% was attained after 10 rounds.


\subsection{Find Naught in Fear for 100 Battles: Empirical Fixation Probabilities}

To assess whether \ip can invade other strategies and resist invasion,
we conducted a large number of simulations of \ip versus other 
strategies for the prisoner's dilemma (Table \ref{table_fixations}).
Starting with $m=1$ player of one type
invading $N-1$ players of a second type, 
we performed simulations as described in Stewart and Plotkin,
i.e. with the exponential imitation dynamic (and $\beta=1$),
donation game matrix, and $N=100$ population size 
\cite{stewart2012extortion} (see Methods). 
In each round of play, every individual plays every other individual, 
moves are randomly flipped with probability $\epsilon=0.05$,
and fitness is computed as total payout versus all other players. 
Note that no tag (type) information was supplied about any player.
Instead, each player was assigned a unique integer ID value, which
other players could use to track their history of play vs. that specific
player, and all players were notified of the ID of any player replaced
by the imitation dynamic (but again given no information about the
type of the new player).

Table \ref{table_fixations} lists the fixation odds ratio
of each strategy versus each other strategy, determined empirically 
via simulation (specifically, it gives the ratio $\rho/\rho_{neutral}$,
where $\rho$ is the observed fixation probability, and $\rho_{neutral}=1/N$
is the fixation probability expected under neutral selection;
hence a table value of 1.0 indicates neutral selection).
In no case was \ip successfully invaded by any other strategy. 
By contrast, \ip is able to invade all other strategies, 
with a fixation probability greater than a neutral mutant 
($\rho>\rho_{neutral}$), 
and in all cases is either the best or second best opponent (i.e. largest 
or second largest value in each column). In the language of the 
Moran process, \ip has higher relative fitness versus all other strategies, 
and as a resident strategy is evolutionarily robust 
(defined as $\rho \le \rho_{neutral}$ for all possible invaders 
\cite{nowak2004emergence}) tested. 
Qualitatively similar results hold for other population sizes 
$N \approx 30$ or greater.
We also simulated with a Moran selection rule, where each round one player is 
selected to reproduce proportionally to fitness and one player is selected to 
be replaced uniformly at random \cite{moran1962statistical} \cite{nowak2006evolutionary}. 
Results were similar. Also, results for simulations using the standard 
prisoner's dilemma score matrix 
(as in \cite{press2012iterated}, instead of the donation game matrix) are 
qualitatively similar. In principle, \ip should excel in any asymmetric game with
similar updating rules (it is not designed specifically for a particular game or
updating rule).


These values reveal much about how \ip competes against other players. \ip is nearly as effective against ALLC as ALLD is, and quickly learns to exploit ALLC, but has a slightly smaller fixation probability because of the information gaining stage. \ip also fares well against ALLD, behaving much like TFT in that it cooperates with other (identified) \ip individuals and defects against ALLD. Outcomes versus ALLD are sensitive to initial population proportion. An invading subpopulation of 10 \ip has an empirically computed fixation probability of $\rho \approx 0.5$ (versus a neutral fixation probability of $\rho = 0.10$).

Versus TFT, \ip does not fall prey to the mismatches in due to errors that TFT is prone to \cite{nowak2006evolutionary}, but may suffer versus TFT in the information gaining period (and so does not fare quite as well as ALLC, but has a higher chance to invade than all other players). Among all strategies in our simulations, \ip is the only strategy to have a fixation probability greater than a neutral mutant ($\rho_{neutral} = 1/N$) versus all other strategies, and the only strategy resistant to invasion by all other strategies.

In general, the ability of \ip to invade other strategies appears to correlate
with its fitness difference vs. those strategies at low population 
fractions (i.e. $m \approx 1$, see Figs. 1 - 2).  For those where
\ip can immediately achieve a strongly positive stationary score difference 
(e.g. vs. ALLC, TFT, \zdr, \zdx), 
it can invade with high fixation probabilities.
For those where \ip is confined to neutral score for low values of $m$
(e.g. vs. ALLD, WSLS), its fixation probabilities are lower.

\begin{table}
\begin{tabular}{c|c|c|c|c|c|c|c|}
& \ip & ALLC & ALLD & TFT & WSLS & \zdr & \zdx\\
\ip &  & 58.10 & 5.50 & 43.60 & 1.96 & 16.30 & 51.01\\
ALLC & 0 & & 0 & 49.48 & 0 & 21.14 & 54.78\\
ALLD & 0 & 59.38 & & 0 & 0.05 & 0 & 0\\
TFT & 0 & 0 & 3.68 & & 0 & 0 & 9.74\\
WSLS & 0 & 34.72 & 0 & 7.11 & & 0.32 & 21.16\\
\zdr & 0 & 0 & 0.86 & 24.07 & 0 & & 27.55\\
\zdx & 0 & 0 & 1.61 & 0 & 0 & 0 &
\end{tabular}
\caption{Fixation odds ratios $\rho/\rho_{neutral}$ of a single row player invading a population of $N-1 = 99$ column players, with an ambient error rate of $\epsilon = 0.05$. At least 10,000 simulations were performed for each pair of types. For \ip, p-values for the null hypothesis of neutral fixation is $p=5 \times 10^{-10}$ for ALLD and $p < 10^{-26}$ otherwise.}
\label{table_fixations}
\end{table}

Smaller values of $\epsilon$ make TFT more challenging to infiltrate, however at $\epsilon = 0.01$ the fixation probability of an \ip mutant is still 8 times the neutral probability. This dependence is due to the relatively large number of rounds needed for TFT to reach its stationary distribution versus some other strategies, and this prolongs the time needed to invade an ambient population of TFT players. \ip is apparently uninvadable by TFT at $\epsilon = 0.01$ and $N=100$, but was invaded once in 10017 simulations for $N=40$.

%
%

\subsection{Robust Zero Determinant Strategies}

Stewart and Plotkin have outlined a series of assumptions under which 
\zdr strategies are robust to all other IPD strategies
\cite{stewart2012extortion}. 
One implicit assumption in this argument is that players' type 
cannot be identified, either by a tag as described by 
Adami and Hintze \cite{adami2013evolutionary} or by statistical 
inference from the history of play as performed by \ip.
As shown in Fig. 2C, \zdr strategies are vulnerable to
invasion by such information players, because the \zdr can
at best guarantee neutral selection 
i.e. ($\overline{S_{I}}-\overline{S_{G}}=0$) vs. the IP invader
at low population fractions ($m \approx 1$), whereas when
the IP invader is in the majority it can gain a strong selective
advantage ($\overline{S_{I}}-\overline{S_{G}} \gg 0$).
In simulations, we found that a tag-based IP (ConSwitch)
invades \zdr at much higher than neutral fixation probability
($\rho/\rho_{neutral} \approx 1.6$, see Fig. 4), and that 
\ip achieved better than neutral invasion success
against \zdr for $\chi \ge 0.8$ even at zero noise ($\epsilon=0$).
We wish to emphasize that our \ip implementation clearly falls
far short of the theoretical IP performance limit as indicated
by ConSwitch.  This mirrors what we saw at low population 
fractions in Fig. 2B, where \ip
falls short of the perfect (neutral) score that ConSwitch attains vs \zdr.
This shortfall is due to the ``cost'' of the infogain phase in the
current IP implementation, which indicates a clear direction
for improvement of our IP implementation.

A second factor that renders \zdr easily prone to invasion by \ip
is the effect of noise.  Even low levels of noise (e.g. $\epsilon=0.01$)
enabled \ip to invade \zdr at better than neutral fixation probability
at all values of $\chi$ (Fig. 4).  
In general, noise appears to degrade the performance of
Markov players such as \zdr even more than it degrades the performance
of \ip.  Specifically, noise reduces \zdr's ability to cooperate
with itself (i.e. the fraction of \zdr-\zdr game outcomes that are CC),
and hence its stationary score (see Fig. 4),
more than it reduces \ip's ability to cooperate with itself
(because its self-recognition algorithm is robust to noise,
and its self-strategy -- ALLC -- is less affected by noise than
\zdr is).



\begin{figure}[h!]
\includegraphics[width=0.4\textwidth]{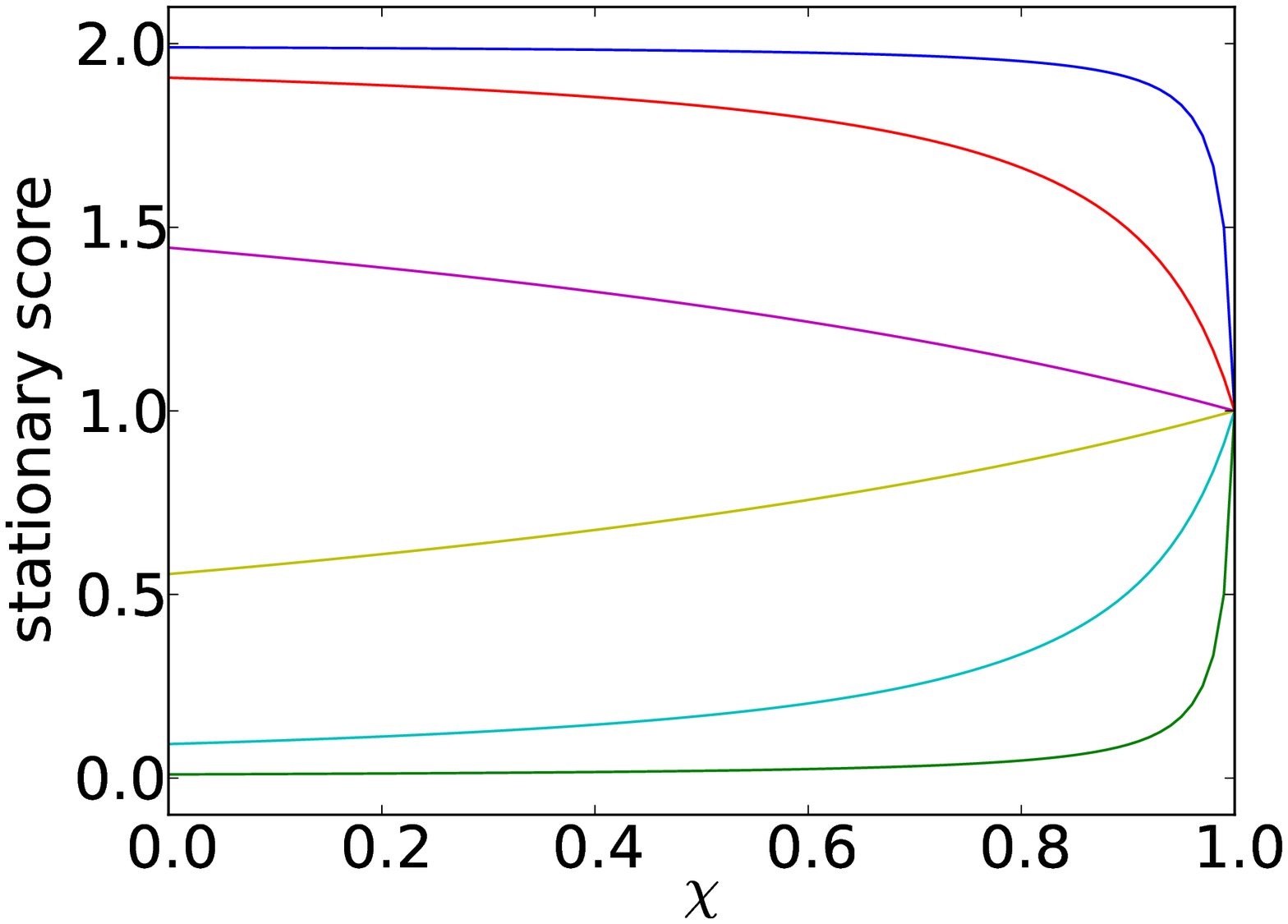}
\includegraphics[width=0.4\textwidth]{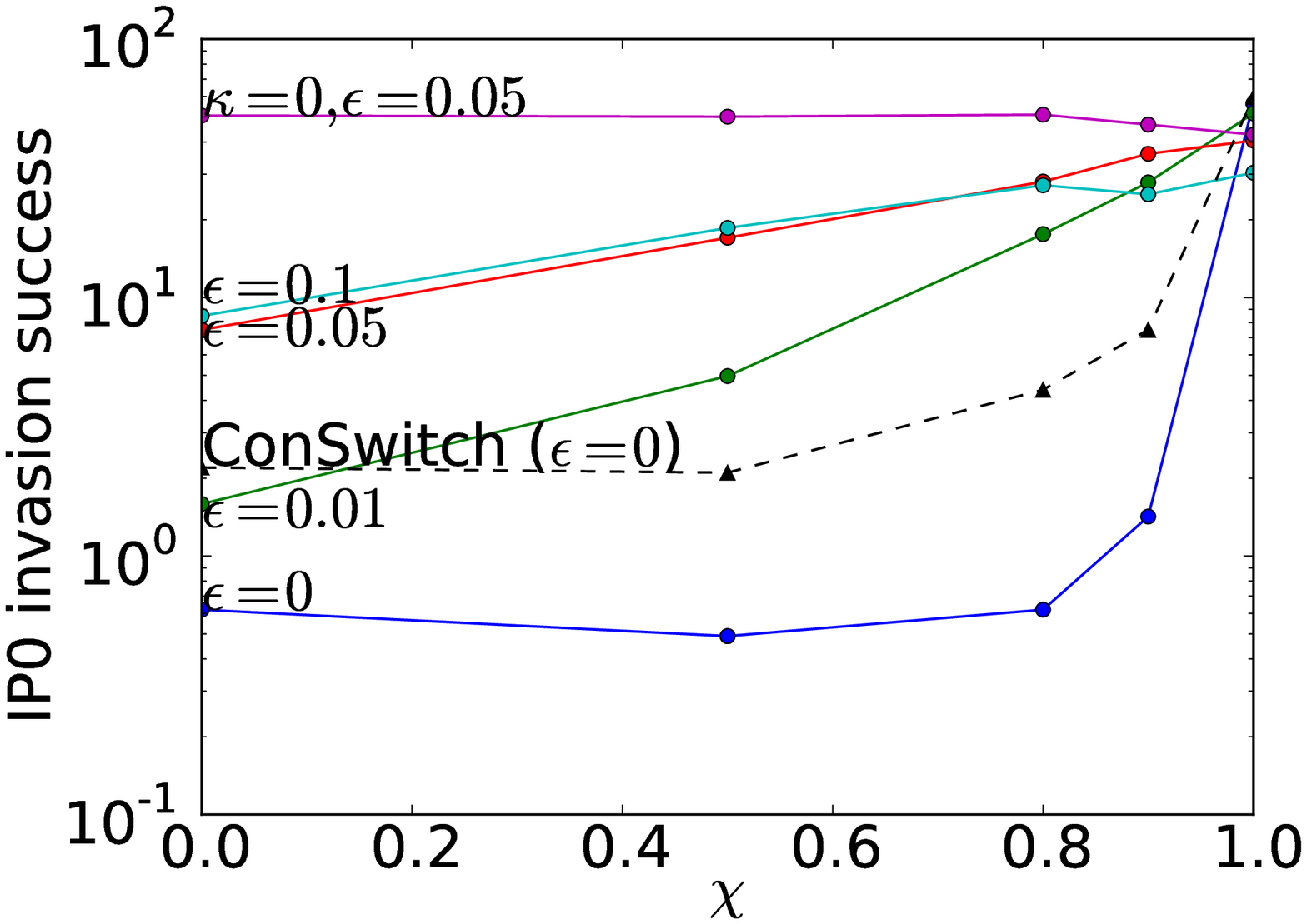}
\caption{Invasion success of IP versus ZD strategies (log-scale, fixation probability of an IP0 invader, normalized so 1.0 = neutral selection) for different levels of noise $\epsilon$. The top plot is extortionate ($\kappa=0$ while the lower three plots have $\kappa = B-C$ so the ZD strategies are cooperative \cite{stewart2013extortion}. As the value of $\chi$ increases, the fixation probability of IP increases. As the amount of noise decreases, the fixation of our implementation of IP approaches the neutral fixation. With no noise, an IP player can empirically invade \zdr at twice the neutral probability (20 out of 1000 simulations with the information phase replaced with tags).}
\label{robustness}
\end{figure}

\section{Discussion}

For zero noise, weak selection, no history, and stationary payouts, the robustness results of Stewart and Plotkin are not contradicted by our empirical results (likewise for the strong selection results in \cite{stewart2014collapse}). Our results, however, indicated that with tagging of player strategies, robust zero determinant strategies can be invaded for non-weak selection and no noise. This should not be surprising from Figures \ref{figure_stationary_score_1} and \ref{figure_stationary_score_2}.  \zdr is not able to generally able invade \ip nor the variety of strategies that \ip is able to invade. For instance, \zdr is neutral versus e.g. ALLC (with $\epsilon=0$), whereas \ip can invade ALLC easily at the same level of noise.
Note that whereas \ip is always able to invade \zdr strategies,
even at zero noise, neither \zdr strategies nor any of the other
strategies we've tested is ever able to invade \ip.

Fixation probabilities for zero-determinant strategies were studied by Stewart and Plotkin \cite{stewart2013extortion} in the case of weak selection. Our results indicate \ip is robust to invasion against all the opponents in Table \ref{table_fixations}. That this generally holds is simply a consequence of the fact that \ip specifically chooses to maximize the stationary score difference with its opponents while obtaining the cooperative payout when playing itself. Therefore, once the information gain phase is over, \ip will fixate at least as well as a neutral mutant strategy, and typically much more often. For \ip to be invaded or resisted better than a neutral mutant, the opponent strategy must somehow exploit the manner in which IP attempts to gain information (perhaps by mimicing \ip to be misidentified as another \ip player), or the information gain phase must be too costly (for instance in a very small population).

We have shown that information strategies utilizing the history of play can be highly effective infiltrators and are very robust against invasion in population games. \ip is able to invade essentially any memory-one strategy for a reasonably large population (close to neutral in the worst cases, vastly dominant in others), with greater success under greater ambient error probabilities (at least in the range we tested, $0 \epsilon 0.1$). We conjecture that for sufficiently large populations \ip is robust to invasion against all memory-one strategies, and also that \ip is neutral or better as an invader of memory-one strategies (with exceptions occurring mainly for small ambient noise and/or weak selection).

While we have discussed our results in the context of the prisoner's dilemma, \ip is effective in principle in \emph{all} population games without significant modification. For any game matrix, \ip will still identify other players' strategies and maximize the difference in stationary payout. Information players should fare well in a variety of other contexts, including asymmetric games and population games on graphs.

We have not attempted to optimize the relative length of the \ip information gain phase, and it is clear in some contexts that finer-tuned play is possible, particularly against generous ZD strategies for the donation game. In particular, very small populations may require a more refined information gain phase. We have also not attempted to optimize against other ``theory of mind'' players that may be able to play effectively against our implementation of an information player. The implementation does attempt to recognize players that are able to change their strategies and respond accordingly, and similar countermeasures could be employed. At some point, for the prisoner's dilemma, play between two players degenerates into an ultimatum game \cite{press2012iterated}, and uncooperative and/or manipulative players could simply be systematically defected against.

\subsection{History of Play}

In \cite{stewart2013extortion}, Stewart and Plotkin argue that (under weak selection, in the absence of ambient noise, and using stationary score as fitness) that one need only consider memory-one strategies in population games to determine evolutionary robustness (extending a similar idea of Press and Dyson for two-player games). Like Adami and Hintze, we find that the history of play allows for highly effective non-memory-one strategies. The assumption that each player has a unique identifier is operationally equivalent to providing in each pairing the history of play of the pair to each player before they select their move, allowing for a personalized response, e.g. the inference of strategy identifying tags. Note carefully that we are not assuming that the history of interactions with other (third-party) players is available to each pair, just the history of play of the specified pairing. Our results show that it is not generally sufficient to only consider memory-one players in population games. 


Another important distinction of \ip is that the individual information players cannot be aggregated as all having the same stationary score with each other. Indeed, the \ip subpopulation is more like a quasispecies with several closely related variants, with each information player potentially identifying a different subpopulation of information players players and having inferred slightly different conditional probability vectors for non-\ip strategies (the information players share no information explicitly). Accordingly, we computed fixation probabilities empirically from large numbers of simulation and cannot rely on the typical analytic formulas for two-type population games (death-birth processes). For larger populations, the deviation from the theoretical values (from the stationary payouts) should be small, since \ip can quickly approach stationary payoffs.

In principle, information strategies could be generalized to detect memory-$m$ players using $m$-rounds of history to form their strategies, such as Tit-for-Two-Tats. This may require a longer information gaining stage, but it is possible to first attempt a memory-one model and use information metrics to determine if an alternate model is required \cite{lee2011empirical}. We leave this topic, and the issue of effective counter-strategies to \ip and other information players, for future work.

\subsection{MaRS}

After circulating an initial draft of this manuscript, the authors were
made aware by Christian Hilbe of a strategy known as MaRS 
(mimicry and relative similarity),
created by Fischer et al \cite{fischer2013fusing}. Like \ip, an individual 
playing the MaRS strategy is an information player and uses a unique 
identifier for each opponent along with the history of
play to formulate a counter-strategy, but unlike \ip uses principles of mimicry 
and imitation to emulate aspects of the classical strategies such as TFT and
WSLS. This leads to the emergence of cooperation by pushing non-cooperative 
opponents toward extinction.

We attempted an implementation of MaRS, which we call \mars to indicate 
potential differences from the implementation in \cite{fischer2013fusing}, and 
simulated population games versus
\ip and the other strategies in Table \ref{table_fixations}. With no noise,
\mars was able to invade TFT (59x neutral), \zdx (35x), \zdr (1.5x) and is 
approx. neutral versus ALLC and WSLS; \ip invades TFT 
at 54x, \zdx at 71x, ALLC at 61x, and ALLD at 7x. Both \mars and \ip are 
robust to invasion by the tested
strategies above, however one additional test strategy GTFT
(generous tit-for-tat) was able to invade \mars 
with a greater than neutral probability ($\approx 600$ out of 10000, 6-fold 
greater than neutral). Interestingly \mars is able to invade GTFT at about
twice the netural probability, while \ip is unable to invade (0.8x) or be 
invaded by GTFT in the same context. This indicates that \mars is 
qualitatively more similar to GTFT than \ip.

At the 5\% noise level, \mars is able to invade ALLC at 46x and \zdx at 3x, and
is invaded by TFT (3x neutral) and \zdr (18x netural). Compare 
to Table \ref{table_fixations} where \ip is robust to invasion and an effective
invader of all the test strategies. Neither \mars nor \ip is able to invade the
other at 0\% or 5\% noise.

Ostensibly the performance of \mars and \ip are somewhat similar for the
prisoner's dilemma in the absence of noise, which can be understood by 
the work of Press and Dyson \cite{press2012iterated}. 
Maximizing the long run (stationary) payoffs of the 
prisoner's dilemma often necessitates some degree of coordination for 
various pairs of opponents by the nature of the payoffs of the game. While \mars 
cooperates with opponents that exhibit a similar history of play by design, \ip 
only cooperates if doing so rewards 
cooperation stationarily, and is not tied in principle to either the prisoner's
dilemma or strategic scenarios that benefit from mimicry or other coordinated
modes of play. For the prisoner's dilemma, the principles underlying \ip
sometimes align with those of \mars, which indicates that \ip too induces the 
emergence of cooperation. But the intentionality is different, since \ip cares
not about cooperation, only in maximizing relative stationary output, 
recapitulating the principle of natural selection.

Nevertheless, it is clear that both \mars and \ip are interesting strategies
that go beyond memory-one dynamics, and that adjustments to the operating rules
of both (such as considerations for noise in the \mars decision rules and 
optimization of the infogain phase of \ip) may result in a further convergence 
of the outcomes for the two strategies for the prisoner's dilemma in the 
contexts tested here. Both \mars 
and \ip are likely to exhibit interesting behaviors for asymmetric and 
higher-dimensional games as the space of strategies beyond memory-one strategies
is explored.

\section{Materials and Methods}

\subsection{Simulations}

Evolutionary simulations were performed using either the Moran process
or the imitation dynamic with selection strength $\sigma=1$ as
in \cite{stewart2012extortion}.  Unless otherwise stated,
all simulations were performed with a total population
size of $N=100$ starting with a single player of the invading type
and run until fixation of either the resident or invading type,
and the donation game score matrix (2, -1, 3, 0) 
as in \cite{stewart2012extortion}.

In most cases, $N_{sim}=10,000$ independent simulations were run for each
(invader, resident) pair, and p-values for the observed number
of successful invasions $k$ were computed under a null hypothesis
$H_0$ assuming a neutral rate of fixation $\theta=1/N$:

\[p_{>} = p(K \ge k|H_0,\theta=1/N) = \sum_{K=k}^{N_{sim}}{{N_{sim} \choose K}\theta^K(1-\theta)^{N_{sim}-K}}\]

Following \cite{press2012iterated} and \cite{stewart2012extortion},
we focus on memory-1 strategies with probability vector
\[ \mathbf{p} = (p_1, p_2, p_3, p_4) = (Pr(C | CC), Pr(C | CD), Pr(C | DC), Pr(C | DD)). \]
Unless otherwise specified in the text, we used the standard
probability vector specified in \cite{stewart2012extortion}
for \zdr (with $\kappa=2,\chi=\frac{1}{2}, \phi=0.1$), and
\zdx (with $\kappa=0,\chi=\frac{1}{2}, \phi=0.1$).

\subsection{Information Player Implementation}

We implemented a basic information player strategy, called \ip, with the following
components: (1) an {\em infogain} phase during
which an \ip player chooses its moves to maximize its information yield
about a new player, both to assess whether it is another \ip
(self vs. non-self), and in the latter case to estimate its strategy
vector; (2) a {\em groupmax} phase during which the IP seeks to
maximize its score relative to the opponent group, by either
cooperating (if the other player is also using ip) or using its
current optimal strategy versus the group (if the other player is not
an IP).  Note that when multiple \ip players are present in a population,
they operate completely independently; they do not share information
or communicate.

\subsection{Basic definitions} \ip records the outcomes of its games vs.
a given player in terms of $(n_{AB},m_{AB})$ pairs, where $A$ is a
possible move (C or D) by itself, $B$ is a possible move by the
other player (C or D), $n_{AB}$ is the total number of times
game outcome $AB$ has occurred with this player, and $m_{AB}$ is the 
number of those cases where the other player's next move was C.
Treating each such case as a binomial event with probability
$p_{AB}=\theta$ (probability of cooperating given game outcome
AB), the posterior distribution is
$p(\theta|n_{AB},m_{AB})=\beta(m_{AB}+1, n_{AB}-m_{AB}+1)$
(i.e. the Beta distribution assuming a uniform prior $p(\theta)=1$), 
the maximum likelihood estimator is
$\hat \theta=m_{AB}/n_{AB}$, and the posterior expectation value
is $\overline{\theta}=E(\theta|n_{AB},m_{AB})=(m_{AB}+1)/(n_{AB}+2)$.
We use the symbol 
$\overline{p}=(\overline{\theta_{CC}},\overline{\theta_{CD}},\overline{\theta_{DC}},\overline{\theta_{DD}})$ to refer to such an infered 
probability vector.

\subsection{Infogain phase} For the first 10 rounds of its play
with another player, \ip chooses its moves
to seek game outcomes $AB$ about which it has the {\em least} information
(smallest number of counts $n_{AB}$).  Specifically, if the current
game outcome was $ab$, then it chooses the move $A$ that minimizes
the expectation value of $n_{AB}$:
\[A_{infogain}=\arg\min_A E(n_{AB}|ab)=\arg\min_A \sum_B{p(B|ab)n_{AB}}\]
where $p(B|ab)=(m_{ab}+1)/(n_{ab}+2)$ for $B=\text{C}$.  In the
case of exact ties (equal expectation values for $A=\text{C,D}$),
the \ip breaks the tie by computing the MD5 hash value of the 
game outcomes history string (e.g. ``CCDC...''), and choosing C
if its least-significant bit is zero, otherwise D.  Note that since
this rule depends only on information
known to both players (their game outcomes), \ip can predict
what moves the other player would choose if it too were an \ip.
(Of course, in the presence of noise $\epsilon$, the confidence
of this prediction drops to $1-\epsilon$).

\subsection{Groupmax phase} During this phase, \ip seeks to maximize
its average relative score vs. the opposing group $G$ (equation \ref{mean_stationary_score}).
Each \ip seeks
to maximize $S_{II}$ (average score versus other \ip players
by cooperating with any player it believes to
be an IP.  With all other players, it applies its current
groupmax probability vector $p_{groupmax}$ chosen to maximize
the difference between the second ($S_{IG}$) and third ($S_{GI}$)
terms above (see below for details).

\subsection{Tag inference} After each game, \ip computes the 
likelihood odds ratio for the observed move $B$ of the other player
assuming either that it is also an \ip, or that is a member
of the opponent group (GP).  This is used to update the total log-odds
ratio for that player:
\[ L'= L + \log{\frac{p(B|\ip,\epsilon)}{p(B|\text{GP})}}\]
where $L$ is the current log-odds ratio, $L'$ is the new
log-odds ratio, and $\epsilon$ is the error rate (frequency at
which a player's moves are flipped).  

During infogain phase,
the move expected from an \ip player is predicted by the infogain
model.  During groupmax phase, it is predicted by a Hidden
Markov Model (HMM) \cite{Rabiner89atutorial} 
consisting of just two states: ALLC (``the other
\ip recognizes me as an \ip, and hence cooperates with me'');
and $p_{groupmax}$ (``the other \ip believes I am not \ip, and 
hence applies $p_{groupmax}$ against me'').  The HMM permits
a transition between either of these states with 1\% probability
per round.  At the beginning of groupmax phase, the prior
probability of the ALLC state is simply set to the current
posterior probability that the other player will classify me as an IP,
specifically $p(\text{ALLC})=1/(e^{-L}+1)$, where $L$ is the
log-odds ratio the other player would compute from my moves.

The conditional probability $p(B|\text{GP})$ is computed according 
to $\overline{p}$, the current
inferred strategy of the opponent.  If \ip has not yet confidently
identified any players as GP (see below for details),
then this $\overline{p}$ is derived solely from the IP player's
game outcomes with this specific player.  Otherwise,
$\overline{p}$ is computed from game outcomes vs.
all GP players that it has confidently identified.
This assumes that all non-\ip opponents use the same strategy and could be
relaxed for games with more than two types.

During infogain phase, an \ip player classifies another player
as confidently GP, based on the p-value of its history of moves
under the null hypothesis that it is an \ip playing infogain moves:
\[p(E \ge e|n,\epsilon) =\sum_{E=e}^n
{{n \choose E}\epsilon^E(1-\epsilon)^{n-E}} \le \alpha\]
where $n$ is the number of games it has played vs. that player,
$e$ is the number of observed mismatches vs. the expected
infogain move (during those games), $E$ is the associated random variable,
and $\epsilon$ is the error rate.  We used
$\alpha=0.01$, for at most one expected false positive
(in a population of at most 100 \ip).  During groupmax phase,
an IP player classifies each player according to its current
log-odds ratio: as an \ip if $L>0$, otherwise as a GP.  
Finally, it estimates the total number
of IPs currently in the population from its posterior
expectation value:
\[ \overline{m}= 1 + \sum_i{p(\ip|L_i)}
= 1 + \sum_i{\frac{1}{e^{-L_i}+1}} \]
where $L_i$ is its log-odds ratio for the hypothesis that 
player $i$ is \ip vs. is a GP
(the one additional count is for the \ip player itself).
When the IP detects birth of a new player, it initializes the new
player's prior log-odds ratio to $L=\log{\frac{\overline{m}}{N}}$.
When it detects the death of a player, if it was confidently
a GP ($L < \log{\alpha}$), that player's outcome counts
$(n_{AB},m_{AB})$ are saved for inclusion in future 
computations of the GP strategy vector $\overline{p}$.

\subsection{Groupmax strategy optimization}
If an \ip is in groupmax phase with at least one player,
it computes an optimal strategy to use against the opposing group,
based on its estimate of the total number of \ip ($\overline{m}$)
and its estimate of the opponent group's strategy vector
($\overline{p}$).  It does this based on seeking the strategy
$p_{groupmax}$ that maximizes the interaction terms of the 
relative score:
\begin{equation}
p_{groupmax}=\arg\max_{q} 
\left[\frac{N-\overline{m}}{N-1}S(q,\overline{p})
-\frac{\overline{m}}{N-1}S(\overline{p},q)\right]
\end{equation}
where $S(p,q)$ is the theoretical long-term score for strategy vector $p$
when playing against strategy vector $q$.  We compute $S(p,q)$
as previously described by \cite{press2012iterated}.  Briefly,
a game between any two players is a Markov chain with states as pairs of plays in each round $\{CC, CD, DC, DD \}$. The chain has a unique stationary distribution $\mathbf{s}$, and the mean of any four-vector $f=(f_1, f_2, f_3, f_4)$ with the stationary distribution for two players $p$ and $q$ is given by the Press and Dyson determinant \cite{press2012iterated}
\begin{equation}
D(p, q, f) =
    \text{det} \left[ \begin{smallmatrix}
                        -1 + p_1 q_1 & -1 + p_1 & -1 + q_1 & f_1 \\
                        p_2 q_3 & -1 + p_2 & q_3 & f_2 \\
                        p_3 q_2 & p_3 & -1 + q_2 & f_3 \\
                        p_4 q_4 & p_4 & q_4 & f_4
                       \end{smallmatrix} \right]
\label{stationary_score}
\end{equation}
when $f$ gives the scores that player $p$ would receive for
outcomes (CC, CD, DC, DD) respectively.  Using this expression, \ip simply searches 
the 4-dimensional strategy vector space by gradient
descent for the $p$ that maximizes the relative score vs.
the opponent strategy $\overline{p}$.

Our implementation of \ip and the simulation code used for this manuscript is available at \url{https://github.com/cjlee112/latude}.

\subsection*{Acknowledgments}

This work was partially supported by the Office of Science (BER), U. S. Department of Energy, Cooperative Agreement No. DE-FC02-02ER63421.

\bibliography{ref}
\bibliographystyle{plain}

\end{document}